\documentclass{cimento}

\usepackage{amsmath,amssymb}
\usepackage{url}
\usepackage{cleveref}
\usepackage{graphicx}
\usepackage{xpunctuate}
\usepackage{siunitx}
\crefformat{figure}{#2figure~#1#3}
\newcommand{\doi}[1]{, DOI:~#1}

\newcommand*{\ie}{i.e\xperiod}

\newcommand*{\Kpipi}{\ensuremath{{K^-\pi^-\pi^+}}\xspace}
\newcommand*{\mKpipi}{\ensuremath{m_{K\pi\pi}}\xspace}
\newcommand*{\tpr}{\ensuremath{{t'}}\xspace}
\newcommand*{\mKpipiTpr}{\ensuremath{(\mKpipi, \tpr)}\xspace} 
\newcommand*{\WaveK}[7]{\ensuremath{{#1}^{{#2}}\,{#3}^{#4}\,\allowbreak{#5}\,{#6}\,{#7}}\xspace}
\newcommand*{\PKSt}{\ensuremath{K^*(892)}\xspace}
\newcommand*{\Prho}{\ensuremath{\rho(770)}\xspace}
\newcommand*{\PKTwoSt}{\ensuremath{K^*_2(1430)}\xspace}
\DeclareSIUnit{\clight}{\text{\ensuremath{c}}}
\DeclareSIUnit[per-mode=symbol]\MeVc{\MeV\per\clight}
\DeclareSIUnit[per-mode=symbol]\GeVc{\GeV\per\clight}
\DeclareSIUnit[per-mode=symbol]\MeVcc{\MeV\per\clight\squared}
\DeclareSIUnit[per-mode=symbol]\GeVcc{\GeV\per\clight\squared}

\definecolor{MpLightGray} {RGB} {211,211,211}
\definecolor{MpFitColor} {RGB} {204,0,9}
\definecolor{MpDarkGray} {RGB} {78,78,78}
\definecolor{MpGray} {RGB} {175,175,175}
\definecolor{MpBlack} {RGB} {0,0,0}
\definecolor{MpWhite} {RGB} {255,255,255}
\definecolor{MpLightRed} {RGB} {255,0,15}
\definecolor{MpRed} {RGB} {204,0,9}
\definecolor{MpLightBrown} {RGB} {197,124,52}
\definecolor{MpBrown} {RGB} {128,64,0}
\definecolor{MpDarkBrown} {RGB} {101,67,33}
\definecolor{MpOrange} {RGB} {228,113,0}
\definecolor{MpLightOrange} {RGB} {255,208,161}
\definecolor{MpHistYellow} {RGB} {255,237,183}
\definecolor{MpHistBrown} {RGB} {78,58,0}
\definecolor{MpLightGreen} {RGB} {82,255,14}
\definecolor{MpGreen} {RGB} {12,128,0}
\definecolor{MpCyan} {RGB} {0,188,227}
\definecolor{MpLightBlue} {RGB} {14,14,210}
\definecolor{MpBlue} {RGB} {8,8,117}
\definecolor{MpVeryLightBlue} {RGB} {127,127,255}
\definecolor{MpHistBlue} {RGB} {179,179,236}
\definecolor{MpDarkBlue} {RGB} {33,13,107}
\definecolor{MpViolet} {RGB} {166,41,194}
\definecolor{MpMagenta} {RGB} {241,52,245}
\definecolor{MpVeryLightRed} {RGB} {243,124,132}

\usepackage{todonotes}

\title{Strange-Meson Spectroscopy with COMPASS}
\author{S.~Wallner\from{ins:x} for the COMPASS Collaboration\thanks{swallner@mpp.mpg.de}}
\instlist{\inst{ins:x} Max Planck Institute for Physics, Germany
}

\setcopyright{S. Wallner on behalf of the COMPASS Collaboration}

\begin{document}

\maketitle

\begin{abstract}
While the spectrum of non-strange light mesons is well known, many predicted strange mesons have not yet been observed, and many potentially observed states require further confirmation.

Using the $K^-$ component of the hadron beam at the M2 beamline at CERN, we study the strange-meson spectrum with the COMPASS experiment. The flagship channel is the \Kpipi final state, for which COMPASS has obtained the world's largest sample. Based on this sample, we have performed the most detailed and comprehensive partial-wave analysis of this final state to date. For example, we observe a clear signal from the well-known \PKTwoSt, and for the first time we study the $K_2(1770)$, $K_2(1820)$, and $K_2(2250)$ in a single analysis. We also find evidence for a supernumerary signal called $K(1630)$, suggesting that this signal is a pseudoscalar exotic strange meson.
\end{abstract}

\section{Introduction}

The PDG~\cite{ref:PDG} currently lists only 25 strange mesons, 9 of which still need confirmation. In addition, many states predicted by quark models are missing.
To establish a complete picture of the light-meson sector, it is important to find all the strange partners of the non-strange light mesons, \ie to complete the corresponding SU(3) flavor nonets. This includes the search for exotic mesons beyond pure quark-model $q \bar q$ states.


At COMPASS, we explore the strange-meson spectrum in diffractive scattering of \SI{190}{\GeVc} negative kaons off a liquid hydrogen target.
Our flagship channel is the \Kpipi final state, which in principle gives access to all strange mesons. COMPASS has acquired the so far world's largest sample of about \num{720000} exclusive events for this final state.
In \cref{sec:pwa} we briefly introduce the partial-wave analysis method used to analyze this sample. In \cref{sec:2p,sec:2m,sec:0m} we show selected results of this analysis. A complete discussion of this analysis can be found in ref.~\cite{Wallner}

\section{The Partial-Wave Analysis Method}
\label{sec:pwa}

We performed a partial-wave analysis in two stages to identify strange mesons appearing in the \Kpipi final state and to measure their masses and widths.

In the first stage, the data is decomposed into contributions from various partial waves.
The partial waves are represented by $a = \WaveK{J}{P}{M}{\varepsilon}{\zeta}{b}{L}$, where $J^PM^\varepsilon$ are the quantum numbers of the \Kpipi system,\footnote{Here, $J$ is the spin of the \Kpipi state and $P$ its parity. The spin projection along the beam axis is expressed in the reflectivity basis~\cite{Chung1975} and given by $M^\varepsilon$.} $\zeta$ is the so-called isobar, \ie the intermediate two-body resonance, and $L$ is the orbital angular momentum between the bachelor particle $b$ and the isobar.
To this end, the distribution in the \Kpipi phase-space variables $\tau$ is modeled as \footnote{The set of partial waves included in $\sum_{a,b}^{\text{waves}}$ is inferred from data based on a large pool of 596 allowed waves using regularization-based model-selection techniques. %
In addition, a so-called flat wave, which has a uniform distribution in $\tau$, is incoherently added to the intensity model. Since the flat wave does not pick up significant intensity, we drop it in \cref{eq:intensity} for simplicity.}
\begin{equation}
\mathcal{I}(\tau; \mKpipi, \tpr) = \sum\limits_{a,b}^{\text{waves}} \psi_a(\tau;\mKpipi) \rho_{ab}(\mKpipi, \tpr)\, \psi_b^*(\tau;\mKpipi). \label{eq:intensity}
\end{equation}
Using the isobar model, we calculate the decay amplitudes $\psi_a$ analytically.
The spin-density matrix $\rho_{ab}(\mKpipi, \tpr)$ represents the production and propagation of the \Kpipi system.
To measure its dependence on the invariant mass \mKpipi of the \Kpipi system and the four-momentum transfer squared \tpr between the beam kaon and the target proton, the decomposition is performed independently in 75 narrow \mKpipi bins and 4 \tpr bins.
No assumption about the resonance content of the \Kpipi system enters at this stage.

In addition to the signal reaction $K^- p \to K^-\pi^-\pi^+ p$, backgrounds from other reactions, e.g\xperiod the production of the $\pi^-\pi^-\pi^+$ final state via diffractive dissociation from misidentified beam pions, leak into the \Kpipi sample. At the first stage, we effectively take these incoherent backgrounds into account by modeling them in terms of \Kpipi partial waves.
Technically, this is done by parameterizing $\rho_{ab}$ as a matrix of rank three~\cite{Wallner}.

In the second stage, we model $\rho_{ab}\mKpipiTpr$ for 14 selected partial waves using physical amplitudes. In this so-called the resonance-model fit we use an incoherent sum over signal and incoherent background contributions.
Each \Kpipi partial wave $a$
is modeled as
\begin{equation}
	\mathcal T_a\mKpipiTpr = \mathcal K(\mKpipi, \tpr) \sum\limits_{k}^{\text{comp.}} \mathcal C_a^k(\tpr) \mathcal D_k(\mKpipi),
\end{equation}
which are coherent sums over various components $k$ that may contribute to wave $a$. These include resonance components for strange mesons and for each wave a so-called non-resonant component that accounts for other coherent production mechanisms leading to the same final state, such as Deck processes.
The dynamic amplitudes $\mathcal D_k(\mKpipi)$ of the resonance components contain relativistic Breit-Wigner amplitudes, whose free parameters are the mass and width of the resonance.
The dynamic amplitudes of the non-resonant components are parameterized by an empirical function.
The coupling amplitude $\mathcal C_a^k(\tpr)$ represents the strength and phase of the component $k$ within the partial wave $a$. We used independent constant coupling amplitudes for each \tpr bin.
The kinematic factor $\mathcal K(\mKpipi, \tpr)$ accounts for the phase-space volume and the production process.

The $\pi^-\pi^-\pi^+$ background, which is the largest incoherent background, is modeled by a fixed parameterization that we determined from the high-precision COMPASS data of the $\pi^-\pi^-\pi^+$ final state~\cite{Adolph2015}.
Other incoherent backgrounds are effectively subsumed by a flexible background parameterization that is fitted to the data.


\section{Partial-Waves with $J^P = 2^+$}
\label{sec:2p}

We included the \WaveK 2+1+\PKSt\pi D and \WaveK 2+1+\Prho K D waves in the resonance-model fit, which both exhibit a clear signal of the well-known \PKTwoSt. The resonance-model describes both waves well. Our determination of the \PKTwoSt mass and width agree with previous experiments with comparably good uncertainties.
The consistent observation of the \PKTwoSt, despite a significant contribution from the $\pi^-\pi^-\pi^+$ background to the \WaveK 2+1+\Prho K D wave, also validates our analysis approach.

\section{Partial-Waves with $J^P = 2^-$}
\label{sec:2m}

\begin{figure}%
	\centering%
	\includegraphics[width=0.95\linewidth]{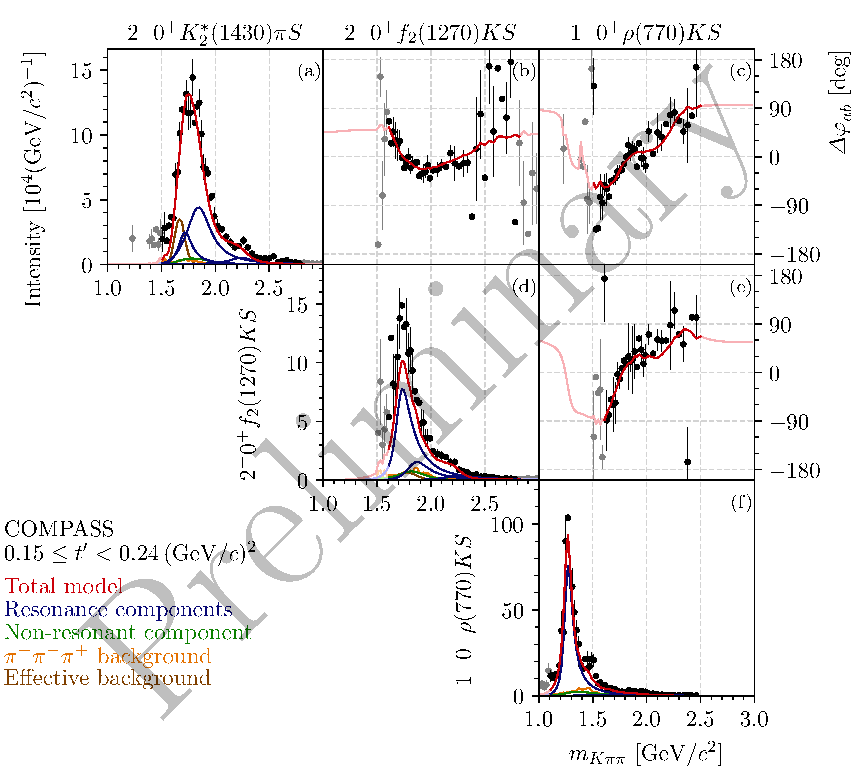}
	\caption{Intensities (diagonal) and relative phases (off-diagonal) of two $2^-$ waves and the \WaveK 1+0+\Prho K S reference wave in an exemplary \tpr bin. The black points represent the measured data. The curves represent the result of the resonance-model fit. The \textcolor{MpRed}{red} curves correspond to the \textcolor{MpRed}{total model}, the \textcolor{MpBlue}{blue} curves to the \textcolor{MpBlue}{resonance components}, the \textcolor{MpGreen}{green} curves to the \textcolor{MpGreen}{non-resonant components}, the \textcolor{MpOrange}{orange} curves to the \textcolor{MpOrange}{$\pi^-\pi^-\pi^+$ background components}, and the \textcolor{MpBrown}{brown} curves to the \textcolor{MpBrown}{effective-background components}. The analytic continuations of all components beyond the \mKpipi fit ranges are shown in lighter colors.}%
	\label{fig:2m}%
\end{figure}

We included four partial waves with $J^P = 2^-$ in the resonance-model fit.
\Cref{fig:2m} shows exemplarily the intensities and relative phases of the \WaveK 2-0+\PKTwoSt\pi S and \WaveK 2-0+{f_2(1270)}KS waves. The intensities shown in panels (a) and (d) exhibit a peak at about \SI{1.8}{\GeVcc} accompanied by a rise of the relative phases shown in panels (c) and (e). Both are well reproduced by an interference between the $K_2(1770)$ and $K_2(1820)$ components.
However, the relative strength of the $K_2(1770)$ and $K_2(1820)$ components is different in the four $2^-$ waves. This can also be seen from the phase difference between the \WaveK 2-0+\PKTwoSt\pi S and \WaveK 2-0+{f_2(1270)}KS waves (b), which is not constant as expected if both waves had exactly the same resonance content.
By simultaneously fitting the four $2^-$ waves, this difference allows to separate the nearby $K_2(1770)$ and $K_2(1820)$.
We exclude the occurrence of only one state in the region around \SI{1.8}{\GeVcc} with a significance of about $11\sigma$.
This is the best evidence for two states so far.

In addition, the intensity spectra exhibit a shoulder at about \SI{2.2}{\GeVcc} accompanied by a rising relative phase. Both are well reproduced by the $K_2(2250)$ component.
For the first time, we studied the $K_2(1770)$, $K_2(1820)$, and $K_2(2250)$ in a single self-consistent analysis. Our determination of masses and widths agree with previous observations.


\section{Partial-Waves with $J^P = 0^-$ and Exotic Strange Mesons}
\label{sec:0m}

In the pseudoscalar sector, the PDG~\cite{ref:PDG} lists the established $K(1460)$ and the unconfirmed $K(1830)$. In addition, the PDG lists a $K(1630)$, whose quantum numbers are actually undetermined. Also, this $K(1630)$ has only been observed by a single experiment so far with a width of only \SI{16}{\MeVcc}, which is unusually small for a strange meson.

\begin{figure}%
	\centering%
	\includegraphics[width=0.45\linewidth]{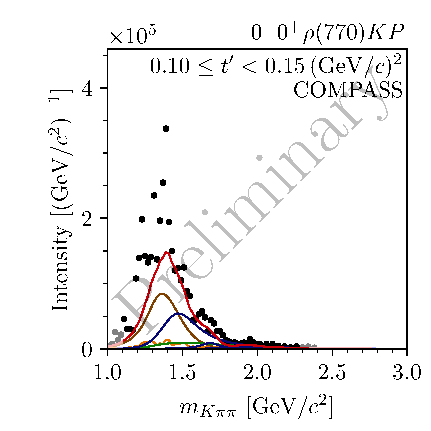}%
	\hspace*{0.05\linewidth}%
	\includegraphics[width=0.45\linewidth]{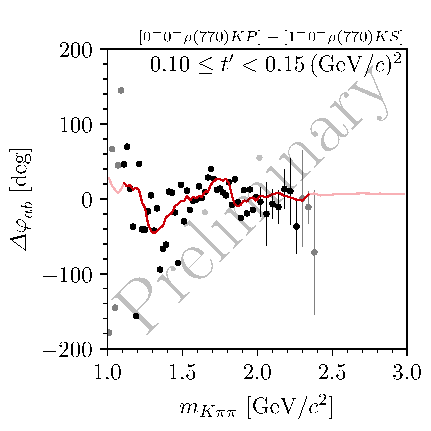}
	\caption{Intensity spectrum (left) and relative phase (right) of the \WaveK 0-0+\Prho K P wave in an exemplary \tpr bin. Same color code as in \cref{fig:2m}.}%
	\label{fig:0m}%
\end{figure}

At COMPASS, we study strange pseudoscalar mesons via their $\rho(770)K$ decay.
The corresponding intensity distribution (\cref{fig:0m}, left) exhibits a peak at about \SI{1.4}{\GeVcc}, which is described mainly by the $K(1460)$ component, as expected, and by the effective-background.
However, this partial wave is affected by known analysis artifacts in the range $\mKpipi \lesssim \SI{1.6}{\GeVcc}$, which are discussed in ref.~\cite{Wallner}. This is keeping us from making robust statements about the $K(1460)$. Hence, we fixed the mass and width of the $K(1460)$ component in the resonance-model fit to values listed by the PDG.

At about \SI{1.7}{\GeVcc}, the intensity distribution exhibits another distinct peak, which is accompanied by a rise of the relative phase of this wave as shown in \cref{fig:0m} (right).
The resonance-model fit yields a significant contribution of the $K(1630)$ component in this region.
While the relative phase is reproduced fairly well, the resonance-model does not perfectly reproduce the intensity spectrum. This imperfection is caused by the mentioned~\cite{Wallner} analysis artifacts.
We verified in detailed Monte Carlo input-output studies and by comparing to well-known states such as the $K_4^*(2045)$, for which we observe similar imperfections in our data (not shown), that the resonance-model fit is not strongly biased by these imperfections. Thus, we can make robust statements about the $K(1630)$ from our data.
We determined the width of the $K(1630)$ to be about \SI{140}{\MeVcc}, which is much broader than observed in the previous measurement.


In total, we observe evidence for three excited pseudoscalar states in a single self-consistent analysis, while quark-model calculations~\cite{Ebert2009} predict only two states in the mass region below $\SI{2.5}{\GeVcc}$. The lightest predicted state agrees best with the $K(1460)$, while the heavier predicted state agrees best with the $K(1830)$.
This suggests $K(1630)$ as a supernumerary resonance-like signal, making it a candidate for an exotic strange meson.
Measurements of further properties of this signal, such as its \tpr dependence, are needed to investigate its nature.
These studies can be carried out partly at COMPASS and at upcoming and planed experiments such as AMBER.

\section{Conclusions}

Many strange mesons claimed in the past require confirmation, and the search for exotic strange mesons has only just begun. Based on the world's largest sample of the $K^-\pi^-\pi^+$ final state, we performed the most detailed and comprehensive analysis of this final state to date. We observe the signal of well-known states such as the \PKTwoSt and the $K_4^*(2045)$. For the first time, we studied the $K_2(1770)$, $K_2(1820)$, and $K_2(2250)$ in a single self-consistent analysis.
In the $J^P = 0^-$ sector we find evidence for a supernumerary resonance-like signal suggesting a pseudoscalar exotic strange meson.

\acknowledgments
The research was funded by the DFG under Germany’s Excellence Strategy - EXC2094 - 390783311 and BMBF Verbundforschung 05P21WOCC1 COMPASS.

\end{document}